\documentclass[twocolumn,showpacs,preprintnumbers,amsmath,amssymb,prl]{revtex4}

\usepackage{graphicx}
\usepackage{dcolumn}
\usepackage{bm}

\begin{document}


\title{Levitation of Bose-Einstein condensates induced by macroscopic
non-adiabatic quantum tunneling}
\author{Katsuhiro Nakamura}
\email{nakamura@a-phys.eng.osaka-cu.ac.jp}
\author{Akihisa Kohi}%
\affiliation{
Department of Applied Physics,
Osaka City University,
Osaka 558-8585, Japan
}

\author{Hisatsugu Yamasaki}
\affiliation{Department of Physics,
Waseda University,
Tokyo 169-8555,
Japan
}%

\author{V\'{\i}ctor M. P\'erez-Garc\'{\i}a}
\email{victor.perezgarcia@uclm.es}
\affiliation{Departamento de Matem\'aticas, E. T. S. Ingenieros Industriales,
Universidad de Castilla-La Mancha, 13071 Ciudad Real,
Spain.
}%
 
\date{\today}

\begin{abstract}
We study the dynamics of two-component Bose-Einstein
condensates  trapped in different vertical positions in the presence of an oscillating magnetic field.
It is shown here how  tuning appropriately the oscillation frequency of the magnetic field
leads to the levitation of the system against gravity. This phenomenon 
is a manifestation of a macroscopic 
non-adiabatic tunneling in a system with internal degrees of freedom.
\end{abstract}


\maketitle


The now classical observation of Bose-Einstein condensation with dilute atomic
vapors in a series of experiments \cite{1} has been the stimulus of
a great number of experimental and theoretical works in the field
of matter waves and macroscopic quantum dynamics.

Shortly after the production of single-component Bose-Einstein condensates (BECs),
multicomponent condensates were observed \cite{4,4b}  in mixtures of 
two hyperfine states of
${}^{87}$Rb, $|F=1,m_{f}=-1\rangle$ and $|F=2,m_{f}=1\rangle$
with and without  ``sag"  corresponding to equal or different trap centers,
respectively. These works, together with other experiments 
which followed soon \cite{5,6,6b} stimulated many studies on
multi-component BECs which open scenarios 
different from those found in single component BECs, for instance
for their ground state and excitations.

In spinor BECs it is easy to induce 
Rabi-type transitions optically. Similar phenomena involving coupling of different states occur
 in tunneling of BECs in optical lattices or double-well potentials \cite{6b,9,Juanjo,10,11,12,Kevre,13,Over}. 
 
In this paper we describe a striking levitation phenomenon which occurs when both the spatial aspects of the dynamics are put together with the two internal degrees of freedom present in a two-component Bose-Einstein condensate in which transitions between components are driven by an oscillating magnetic field. In this scenario the dynamics splits into a fast complex spatio-temporal oscillation of the condensate wavefunctions together with a slow dynamical levitation of the total center of mass against gravity with neither applied mechanical force nor associated classical trajectories.

\begin{figure}[!htb]
\includegraphics[width=0.85\columnwidth]{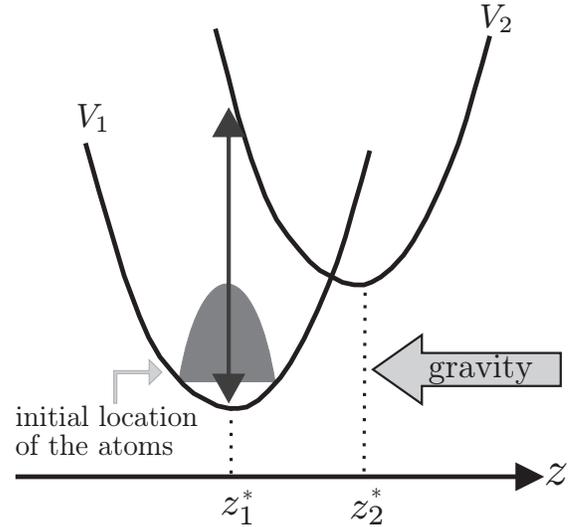}
\caption{Schematic plot of the setup studied in this paper. Each of the components of a
 two-component BEC is subject to 
the action of gravity and different confining potentials $V_1$ and $V_2$ with minima $z_1^{*}$ and $z_2^{*}$, respectively. An oscillating magnetic field induces transitions between both states. A Franck-Condon type vertical transition is indicated by the black arrow. Initially all atoms are in the first component.}
\label{fig1}
\end{figure}

In this paper we will study the dynamics of two-component BECs where 
 each component is trapped in different vertical positions as schematically indicated in Fig. \ref{fig1}.
An optical dipole trap together with a magnetic field gradient can provide such kind of confinement \cite{11,12}.
To fix ideas we will consider our system to be the $|F=1,m_{f}=-1\rangle$ and $|F=2,m_{f}=1\rangle$
hyperfine states of ${}^{87}$Rb coupled by an oscillating magnetic field of frequency $\Omega$ inducing transitions between them. We shall pay a special attention to the situation 
of the Franck-Condon-type vertical transition \cite{13,14} shown in Fig.1. 
To simplify the treatment yet preserving the 
spatial aspects of the dynamics we will assume our system to be magnetically tightly 
confined along one of the transverse 
directions \cite{PG98} to two effective dimensions.

The Hamiltonian density of our system is 
\begin{equation}
\begin{split}
\mathcal{H} =& \sum_{j=1}^{2}\left(\frac{\hbar^{2}}{2m}|\nabla \psi_{j}|^{2}
+V_{j}|\psi_{j}|^{2}+\frac{U_{jj}}{2}|\psi_{j}|^{4}\right) \\
                     & +U_{12}|\psi_{1}|^{2}|\psi_{2}|^{2}+B\cos(\Omega
t)(\psi_{1}^{*} \psi_{2}
+\psi_{2}^{*} \psi_{1}),
\end{split}
\label{eq:2}
\end{equation}
where $V_j$ are the confining potentials acting on each of the species given by 
\begin{equation}\label{eq:1}
V_j(x,z)=\frac{1}{2}m\omega^2\left[x^2+(z-z_j)^2\right]+\epsilon_j + mgz,
\end{equation}
where $\epsilon_j$ stands for the internal electronic energy and $g$ is the
gravity constant. $V_j(x,z)$ has the minimum value
$mz_jg-\frac{m}{2}(\frac{g}{\omega})^2+\epsilon_j$
at $(x,z)=\left(0,z_j^{*}=z_j-g/\omega^2\right)$.
$U_{11},U_{22}$ and $U_{12}=U_{21}$ are the effective nonlinear interaction coefficients 
defined by
$U_{ij}=4\pi \hbar^2 a_{ij}N/m\ell$ in
2-dimensional (2D) space
where $N$ is the total number of atoms and $a_{ij}$ are the scattering lengths for binary collisions. Finally, 
$\ell=\sqrt{\hbar/m\omega}$ is a characteristic length.

 The Gross-Pitaevskii equation (GPE) for the mean field dynamics of our system is derived 
through Lagrange equations with use of  the Hamiltonian $H=\iint \mathcal{H} dxdz$.
Additionally, 
we change to new variables scaled as $
\omega t\rightarrow t, \, x/\ell \rightarrow x, \,
z/\ell \rightarrow z,
\, \ell\psi \rightarrow \psi$ and obtain
\begin{subequations}
\label{eq:6}
\begin{eqnarray}
i \frac{\partial \psi_1}{\partial
t}=&&\left[-\frac{1}{2}\nabla^{2}+V_1^{\prime}+U_{11}^{\prime}|\psi
_1|^2+
U_{12}^{\prime}|\psi_2|^2\right]\psi_1 \nonumber\\
&& + B^{\prime}\cos(\Omega^{\prime} t)\psi_2 \label{eq:6a}\\
i \frac{\partial \psi_2}{\partial
t}=&&\left[-\frac{1}{2}\nabla^{2}+V_2^{\prime}+U_{22}^{\prime}|\psi
_2|^2+
U_{21}^{\prime}|\psi_1|^2\right]\psi_2 \nonumber \\
&& + B^{\prime}\cos(\Omega^{\prime} t)\psi_1
\label{eq:6b}
\end{eqnarray}
\end{subequations}
where $V_{j}^{\prime}\equiv V_{j}^{\prime}(x,z) =
\frac{1}{2}[x^2+(z-z_j)^2]+g^{\prime}z +\epsilon _j^{\prime} $
with  $U_{ij}^{\prime }=4\pi N a_{ij}/\ell, B^{\prime}=B/\hbar \omega,
\Omega^{\prime}=\Omega/\hbar \omega,
g^{\prime}=g/\ell\omega^{2}$ and
$\epsilon _j^{\prime}=\epsilon_j/\hbar \omega$
being  dimensionless constants. The normalization condition for $\psi$ is 
$\iint(|\psi_1|^2+|\psi_2|^2)\; dx dz=1$.

\begin{figure}[!htb]
\includegraphics[width=\linewidth]{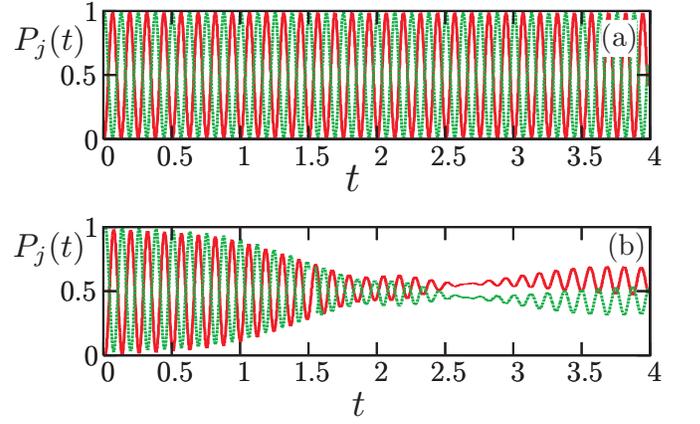}
\caption{[Color online] Time dependence of the population mixing for the case of resonance of Franck-Condon
type. Dashed and solid lines stand for $P_1 = \iint |\psi_1|^2 $ and $P_2 = \iint |\psi_2|^2 $,
respectively for $B^{\prime}=50$.
(a) $\delta z=1.0$ (b) $\delta z=5.0$.}
\label{fig2}
\end{figure}

\begin{figure}[!htb]
           \includegraphics[width=\columnwidth]{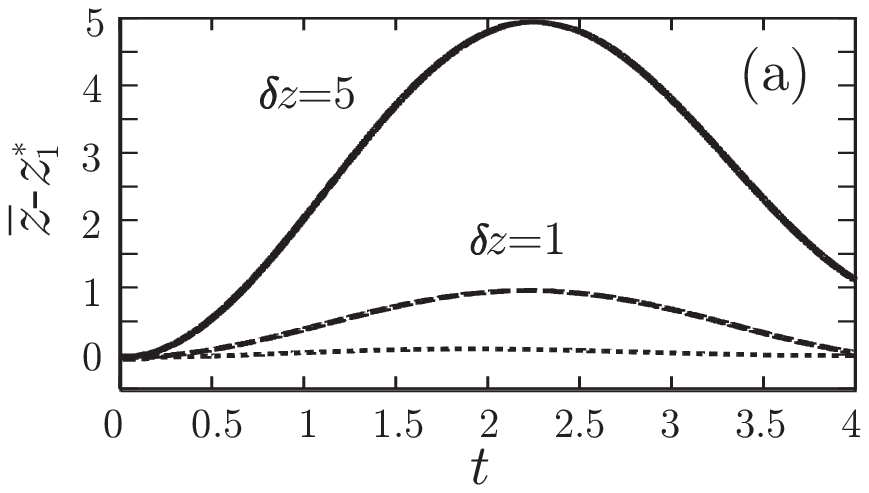}
            \includegraphics[width=\columnwidth]{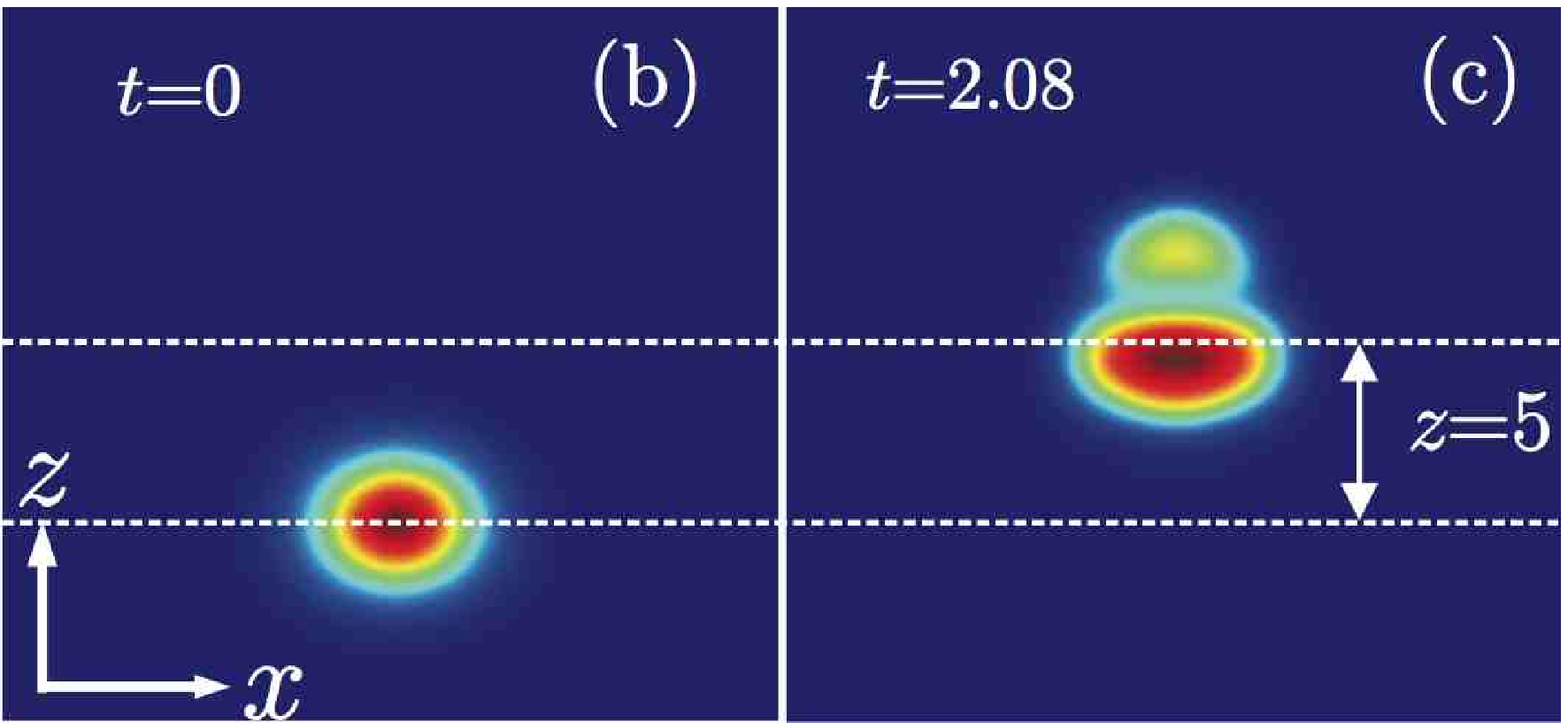}
          \caption{(a) Time evolution of the center of mass 
          $ \bar{z}=\int \bm{\psi}^{\dagger}z\bm{\psi}d^2\bm{r}=\int z \left(|\psi_1|^2+|\psi_2|^2\right)d^2\bm{r}$ for $B^{\prime}=50$ and $\delta z=1.0$ (broken line)  and $\delta z=5.0$ (solid line).
The dotted line stands for the case far 
 from the standard resonance ($\Omega'=1, \delta \epsilon' = 500$) with $\delta z= 5$. (b) and (c) Pseudocolor plots of $|\Psi(x,z,t)|^2 = |\psi_1(x,z,t)|^2 + |\psi_2(x,z,t)|^2$ for $t=0$ and $t=2.08$ on the region $(x,z) \in [-6.5,6.5]\times [-9.6,9.6]$ showing the rising (white broken lines) of the total wavefunction.}
\label{fig3}
\end{figure}

Initially we prepare
a circularly-symmetric Gaussian wavepacket with its center of mass
at ${\bf r}_1=(0,z_1^{*})$,
\begin{equation}
\psi_1=\frac{1}{ \sqrt{\pi}
          \left( 1+\frac{U_{11}^\prime}{2\pi} \right)^{1/4}}
\exp \left(-\frac{x^{2}+(z-z_1^{*})^2}{2
\sqrt{1+\frac{U_{11}^{\prime
}}{2\pi}}}\right)
\label{eq:7}
\end{equation}
which minimizes the energy
$E=\int d^{2} {\bf r}( \frac{1}{2}|\nabla
\psi_1|^{2} + V_{1}^{\prime}|\psi_1|^{2}
         + \frac{U_{11}^{\prime}}{2}|\psi_1|^{4} )$
  over a family of gaussian functions
and approximates the ground state when all atoms are in state $|1\rangle$. 
We then apply an external 
microwave field that induces a Franck-Condon
type transition between the species with
frequency $\Omega^{\prime}=\Delta V_{12}^{\prime}=
V_2^{\prime}(0,z_1^{*})-V_1^{\prime}(0,z_1^{*})$.
Without loss of generality, we will consider the case $z_2^* = -z_1^*$ .
After the oscillating magnetic field is 
switched on, we study the dynamics according to Eq. (\ref{eq:6}).

\begin{figure*}[!htb]
\includegraphics[width=2.1\columnwidth]{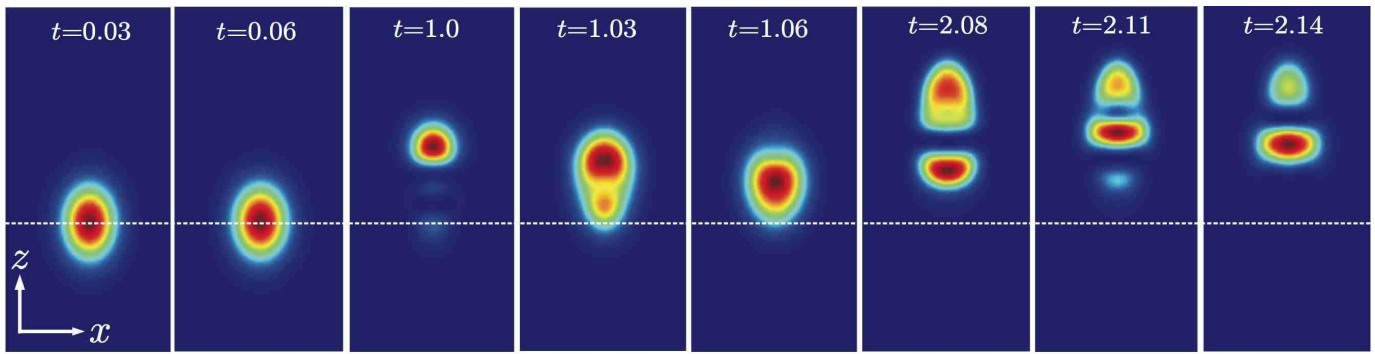}
\caption{Pseudocolor plots of $|\psi_2(x,z,t)|^2$ on the region $(x,z) \in [-6.5,6.5] \times [-7.1, 6]$ for different values of $t$ and $\delta z=5.0$, $B^{\prime}=50$. The time spacing between all pannels, except for pannels 2 and 3 and pannels 5 and 6, is half the Rabi period $\tau_{RB}$.}
\label{fig4}
\end{figure*}

\begin{figure}[!htb]
\includegraphics[width=\columnwidth]{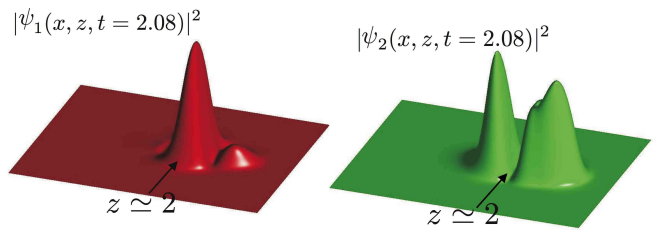}
\caption{Surface plots of $|\psi_1(x,z,t=2.08)|^2$ (red) and $|\psi_2(x,z,t=2.08)|^2$ (green) for the situation described in Fig. \ref{fig4}. The arrow indicates the location of maximum density of $|\psi_1|^2$ corresponding to a low-density region of $|\psi_2|^2$. The spatial region shown in the plot is $(x,z) \in [-6.5,6.5] \times [-6.5, 6.5]$}
\label{figultima}
\end{figure}

Since our results are based on long-time numerical simulations of Eqs. (\ref{eq:6}) we have integrated them
by different numerical methods: an alternating direction implicit method,
a Crank-Nicholson's method, and a split-step method with pseudospectral integration of the spatial derivatives \cite{18}, all of which lead to identical results.
The values of the scaled parameters used throughout the text are: $U_{11}^{\prime}=100$, $U_{22}^{\prime}=97$, $U_{12}^{\prime}=94$;
$g^{\prime}=0.1$; $z_1^{*}=-2.5$ and $z_2^{*}=-z_1^{*}$ with
$\delta z=z_2^{*}-z_1^{*}= 1,3$ and $5$.
$\epsilon_1^{\prime}=-250$, $\epsilon_2^{\prime}=250$ and
$\delta\epsilon^{\prime}=\epsilon_2^{\prime}-\epsilon_1^{\prime}=500$ \cite{20}.

First we study the population mixing between the components.
In the standard resonance case ($\Omega^{\prime}=\delta\epsilon^{\prime}=
\epsilon_2^{\prime}-\epsilon_1^{\prime}$) without
coupling with
the orbital degrees of freedom,
the Rabi formula implies that each $P_j(t)$($=\int|\psi_j|^2d^2{\bf r}$)
 shows a fast oscillation
whose frequency is proportional to the magnetic field amplitude
$B^{\prime}$ \cite{19}.
In our Franck-Condon type transition ($\Omega^{\prime}=\Delta
V_{12}^{\prime}$), the spatial orbital functions of BEC are coupled with the electronic degrees of
freedom, and the Rabi oscillation (with period $\tau_{RB}=2\pi/B^{\prime}$) becomes modulated sooner or later.
In Fig. \ref{fig2} we show the time dependence of both populations
for different values of the distance between trap centers $\delta z$.  We find that 
unless $\delta z$ is sufficiently small,
the regular oscillation responsible for population mixing becomes suppressed as time elapses, 
which is attributed
to the energy conversion from electronic to mechanical energy.

We have also studied the motion of the total center of mass (CM) of the system 
along the vertical direction defined by 
$ \bar{z}=\int
\bm{\psi}^{\dagger}z\bm{\psi}d^2\bm{r}=\int z \left(|\psi_1|^2+|\psi_2|^2\right)d^2\bm{r}.$

In Fig. \ref{fig3}(a), except for the lowest curve, we show typical examples of the time evolution of the center of mass. In the Franck-Condon resonance situation described above we find a rising of the
total center of mass against gravity. This is an interesting phenomenon which happens 
while the underlying population mixing shows a fast Rabi oscillation. In our numerical simulations we observe 
that the CM of the condensate exhibits a smooth and upward acceleration,
reaching a maximum height $\sim z_2^{*}$ for $t = 2.2$ corresponding to 
($2.2/\tau_{RB} \sim 18$ Rabi periods) for $B^{\prime}=50$.
This physical phenomenon is a \emph{levitation of the condensate through a 
macroscopic quantum transition which is not accompanied by neither an applied
mechanical force
nor a  classical trajectory}. Figs. \ref{fig3}(b) and (c) are the total probablility density $|\Psi(x,z,t)|^2 = |\psi_1(x,z,t)|^2 + |\psi_2(x,z,t)|^2$  before and after the levitation, respectively. The levitation effect becomes more pronounced
as $\delta z$ is increased. 
It is very interesting to point our that no levitation occurs in the 
case far from the standard resonance ($\Omega' \ll \delta \epsilon'$), which is confirmed in 
the lowest curve in Fig. \ref{fig3}(a). 

Fig. \ref{fig4} shows the spatiotemporal dynamics of the wavefunction $|\psi_2(x,z,t)|^2$
 for times located in the three regions which are clearly discriminated in the dynamics for $\delta z=5$. 
 First, in the early stage region ($t=0\sim 0.5$) the
wavefunction oscillates near the minimum ($z_1^{*}$) of the lower potential well.
As time goes on ($t=1\sim 1.5$), the central position of the solutions moves
towards the minimum of the upper well ($z_2^{*}$). Finally,
when the CM position reaches its peak value $\sim z_2^{*}$ for $t=2\sim2.2$, the wavefunction 
has a double-humped structure with its valley located near the minimum of the upper well. 
Fig. \ref{figultima} shows that the two components $|\psi_{1,2}|^2$ form 
a domain structure in this time region, a typical feature of the system
 due to the nonlinear interactions. The nonlinearity plays an important role in the formation of
 domain structures during the transitions, a feature which is not present in the noninteracting case.

The origin of this transport phenomenon is that the condensate is effectively driven 
by the oscillating magnetic field to the higher energy
states of the upper level, thus leading to a Franck-Condon type vertical transition. 
After the transition the new position is unstable, and
 the atoms relax towards the new minimum (at the position with larger $z$), leading to the observed transport. 

\begin{figure}[!htb]
\includegraphics[width=\columnwidth]{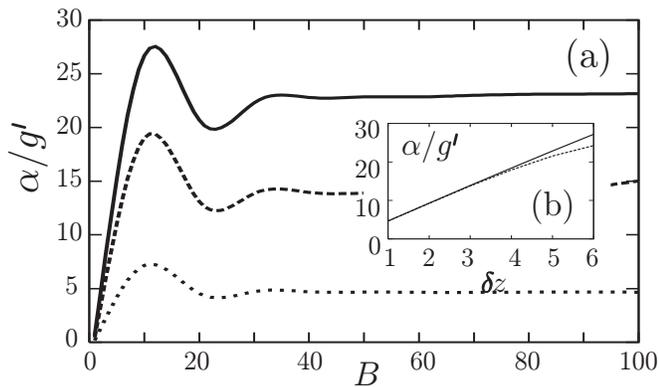}
\caption{ Upward acceleration rate ($\alpha$) for the motion of the center of mass during the levitation process
 (a) $B^{\prime}$ dependence in the cases $\delta z=1$(dotted line), 
$\delta z=3$ (dashed line), $\delta z= 5$ (solid line); (b)
 Dependence of $\alpha$ on $\delta z$ for $B^{\prime}=50$ the Franck-Condon (solid line) and standard (dotted line) resonances.}
\label{fig5}
\end{figure}

We have studied the dependence of the levitation phenomenon on the problem parameters $B^{\prime}$  and $\delta z=z_2^{*}-z_1^{*}$. In Fig. \ref{fig5}(a) we show the 
dependence on $B'$ of the 
upward acceleration rate, $\alpha$, evaluated by using a quadratic function of $t$
(:$\bar{z}=z_1^* +\frac{1}{2}\alpha t^2$) which is fitted to the CM data in the early stage region. 
Interestingly, $\alpha$ has a maximum at some optimal value of $B^{\prime}$.
This fact indicates that the Rabi oscillation is essential for the energy conversion
between electronic and orbital degrees of freedom; however, when the Rabi 
oscillation is too fast to guarantee the relaxation towards $z_2^{*}$, there is a reduction of the effective energy conversion.

In Fig. \ref{fig5}(b) we see that under a fixed $B^{\prime}$, $\alpha$ increases 
monotonically in $\delta z$,
but indicates a slight derivation from the linear law in the case of
non-Franck-Condon type transition (e. g. in the case of standard resonance, $\Omega' = \delta \epsilon'$).

In conclusion, we have studied the dynamics of two-component Bose-Einstein
condensates subject to an a.c. driving magnetic field. 
In the case of Franck-Condon's vertical transition in which the
driving frequency
$\Omega$ corresponds to the energy difference between the
trapping potentials $V_1$ and $V_2$ at the minimum position $z_1^{*}$ of $V_1$, the condensate initially located around $z_1^{*}$ is
effectively driven to unstable states of $V_2$, relaxing
towards the potential minimum $z_2^{*} (=z_1^{*}+\delta z)$ of $V_2$.
In this situation the population mixing shows a fast Rabi oscillation, while
the mean center-of-mass of the condensate exhibits a smooth and upward acceleration
with its rate ($\alpha$) showing a maximum at some optimal magnetic-field 
amplitude and a monotonic increase in $\delta z$. This phenomenon is a manifestation of macroscopic
non-adiabatic quantum tunneling with internal degrees of freedom, which differs essentially from 
other quantum tunneling processes with Bose-Einstein condensates discussed up to now.


This work has been partially supported by grants
BFM2003-02832  (Ministerio de Educaci\'on y
Ciencia, Spain) and PAI-05-001 (Consejer\'{\i}a de Educaci\'on y
Ciencia de la Junta de Comunidades de Castilla-La Mancha, Spain).


\begin{thebibliography}{99}
\bibitem{1} M. H. Anderson, \emph{et al.}, Science {\bf 269}, 198 (1995); K. B. Davis, \emph{et al.}, Phys. Rev. Lett. {\bf 75}, 3969 (1995); C. C. Bradley, C. A. Sackett, and R. G. Hulet, Phys. Rev. Lett. {\bf 78}, 985 (1997).
\bibitem{4} D. Hall, \emph{et al.}, Phys. Rev. Lett. {\bf 81},1539 (1998).
\bibitem{4b} D. Hall, \emph{et al.}, Phys. Rev. Lett. {\bf 81},1543 (1998).
\bibitem{5} M. R. Matthews {\it et al.}, Phys. Rev. Lett. {\bf 81}, 243 (1998).
\bibitem{6} J. Stenger, \emph{et al.}, Nature (London) {\bf 396} 345 (1998); H.-J. Miesner, \emph{et al.}, Phys. Rev. Lett. {\bf 82}, 2228 (1999).
\bibitem{6b} M.R. Matthews, \emph{et al.}, Phys. Rev. Lett. \textbf{83},  2498 (1999).
\bibitem{9} J. Williams, \emph{et al.}, Phys. Rev. A. {\bf 59}, R31 (1999);
M. R. Matthews, \emph{et al.},  Phys. Rev. Lett. {\bf 83}, 3358 (1999).
\bibitem{Juanjo}{J. J. Garc\'{\i}a-Ripoll, V. M. P\'erez-Garc\'{\i}a, F. Sols, Phys. Rev. A \textbf{66}, 021602(R) (2002).}
\bibitem{10} B. Wu and Q. Niu, Phys. Rev. A. {\bf 61}, 023402 (2000);
                O. Zobay and B. M. Garraway, Phys. Rev. A. {\bf 61}, 033603 (2000);
                J. Liu, B. Wu, Q. Niu, Phys. Rev. Lett.{\bf 90}, 170404( 2003);
                E. A. Ostrovskaya and Y. S. Kivshar, Phys. Rev. Lett. {\bf
                92}, 180405 (2004).
\bibitem{11} M. Cristiani, \emph{et al.}, Phys. Rev. A. {\bf 65}, 063612 (2002);
A. Smerzi, \emph{et al.}, 
Phys. Rev. Lett. {\bf 79}, 4950 (1997).
\bibitem{12} M. D. Barrett, J. Sauer and M. S. Chapman,
Phys. Rev. Lett. {\bf 87}, 010404 (2001);
Y. Shin, \emph{et al.}, Phys. Rev. Lett. {\bf 92},150401 (2004);
M. S. Chang, \emph{et al.}, Phys. Rev. Lett. {\bf 92},140403 (2004).
\bibitem{Kevre}{B. Deconinck, P. G. Kevrekidis, H. E. Nistazakis, D. J. Frantzeskakis, Phys. Rev. A \textbf{70}, 063605 (2004);  P. G. Kevrekidis,  \emph{et al.}, Phys. Rev. E \textbf{72} 066604 (2005).}
\bibitem{Over}{M. Albiez, \emph{et al.}, Phys. Rev. Lett. 95, 010402 (2005); R. Gati, \emph{et al.}, 
Phys. Rev. Lett. 96, 130404 (2006).}
\bibitem{13}  L. D. Landau and E. M. Lifshitz, {\it  Quantum
Mechanics} (Pergamon, Oxford, 1958).
\bibitem{14}   J. Franck, Trans. Faraday Soc. 21, 536 (1925); E. U. Condon, Phys. Rev. \textbf{28}, 1182 (1926); \emph{ibid} \textbf{32}, 858 (1928).
\bibitem{PG98} V. M. P\'erez-Garc\'{\i}a, H. Michinel, H. Herrero, Phys. Rev. A \textbf{57}, 3837 (1998).
\bibitem{18} See, e. g., S. K. Adhikari, Phys. Rev. E. {\bf 63} 056704 (2001); V. M. P\'erez-Garc\'{\i}a, X. Liu, Appl. Math. Comput. \textbf{144} (2003) 215.
\bibitem{20}  Although $\epsilon_j^{\prime}$ is of the order of
GHz/Hz our results should be valid for any 
$\delta\epsilon^{\prime}$, provided $\delta\epsilon^{\prime} \gg 1$
 as it happens for Franck-Condon or standard 
resonances.
This is so because in the rotating-wave approximation (RWA), the unitary transformation of wave 
functions that makes static the oscillating field produces in the diagonal 
term the difference between
$\Omega^{\prime}$ and $\delta\epsilon^{\prime}$.
This difference is much less than unity in the case of Franck-Condon 
resonance and vanishes in the case of standard resonance.
Since we are not working in the RWA, we use the oscillating field as it 
stands and choose a large enough value of $\delta\epsilon^{\prime}=500$.
\bibitem{19} J. J. Sakurai, {\it Modern Quantum Mechanics} (Addison Wesley,
New York, 1994).

\end{thebibliography}
\end{document}